\documentclass[aps, prd, twocolumn, groupedaddress, showpacs, showkeys]{revtex4}
\usepackage{graphicx}
\usepackage{dcolumn}

\begin{document}

\newcommand{\Ob}{\ensuremath{\Omega_{\rm b}}}
\newcommand{\Obhh}{\ensuremath{\Omega_{\rm b}h^{2}}}
\newcommand{\Oc}{\ensuremath{\Omega_{\rm c}}}
\newcommand{\Ochh}{\ensuremath{\Omega_{\rm c}h^{2}}}
\newcommand{\Om}{\ensuremath{\Omega_{\rm m}}}
\newcommand{\Omhh}{\ensuremath{\Omega_{\rm m}h^{2}}}
\newcommand{\As}{\ensuremath{A_{\rm s}^{2}}}
\newcommand{\At}{\ensuremath{A_{\rm t}^{2}}}
\newcommand{\Ad}{\ensuremath{A_{\rm d}^{2}}}
\newcommand{\ns}{\ensuremath{n_{\rm s}}}
\newcommand{\Power}[1]{\ensuremath{{\cal P}_{\rm #1}}}
\newcommand{\fd}{\ensuremath{f_{\rm d}}}
\newcommand{\ft}{\ensuremath{f_{\rm t}}}
\newcommand{\zr}{\ensuremath{z_{\rm r}}}

\title{WMAP constraints on inflationary models with global defects}

\author{Neil Bevis} 
\email{n.a.bevis@sussex.ac.uk}
\author{Mark Hindmarsh} 
\email{m.b.hindmarsh@sussex.ac.uk}
\author{Martin Kunz} 
\email{m.kunz@sussex.ac.uk}
\affiliation{Astronomy Centre, Department of Physics \&
Astronomy, University of Sussex, Brighton, BN1 9QH, United Kingdom}

\date{15 July 2004}

\begin{abstract}
We use the cosmic microwave background angular power spectra to place
upper limits on the degree to which global defects may have aided
cosmic structure formation. We explore this under the inflationary
paradigm, but with the addition of textures resulting from the
breaking of a global $O(4)$ symmetry during the early stages of the
Universe. As a measure of their contribution, we use the fraction of
the temperature power spectrum that is attributed to the defects at a
multipole of 10. However, we find a parameter degeneracy enabling a
fit to the first-year WMAP data to be made even with a significant
defect fraction. This degeneracy involves the baryon fraction and the
Hubble constant, plus the normalization and tilt of the primordial
power spectrum. Hence, constraints on these cosmological
parameters are weakened. Combining the WMAP data with a
constraint on the physical baryon fraction from big bang
nucleosynthesis calculations and high-redshift deuterium abundance
limits the extent of the degeneracy and gives an upper bound on the
defect fraction of 0.13 (95\% confidence).
\end{abstract}


\maketitle

\section{Introduction}\label{sec:intro}

Recent measurements of the cosmic microwave background (CMB), notably
the first year WMAP data \cite{Hinshaw2003, Kogut2003}, have proved
highly successful in probing the early stages of cosmic structure
formation. The observed CMB anisotropies may be produced by taking
adiabatic primordial perturbations of roughly the Harrison-Zel'dovich
form and evolving these using well-understood, linear
physics. Further, the parameter values that are
required for this process to a give a match to the data are consistent
with those measured using other astronomical techniques. That the
primordial power spectrum predicted by many models of inflation is of the
required form has become an important success of the inflationary paradigm.

On the other hand, a less attractive property of the paradigm is that
successful inflationary models may involve quite different fields,
interactions and levels of physical motivation. Here we address the
issue using CMB power spectra to constrain models of hybrid inflation
\cite{Linde1994,Lyth1999} that involve the formation of
topological defects as inflation ends \cite{Copeland1994}. Such
models, however, do not fit exactly into the above regime. With the
existence of topological defects, the seeding of cosmic structure
continues after inflation ends, for the defects further perturb the
cosmic fluid as long as they continue to be present. For a detailed
review of structure formation with defects see Ref.
\cite{Durrer2002}, but generally, a defect-dominated temperature power
spectrum does not have pronounced acoustic peaks \cite{Albrecht1996}.
Hence, if defects are added to a passive-evolution case and the
normalization reduced to maintain the fit to data on large scales,
then the acoustic peaks are slightly suppressed. That is, however,
assuming that the other cosmological parameters are not also changed.

Earlier work along these lines 
\cite{Battye2000, Bouchet2001, Contaldi1999, Jeannerot1997, Pogosian2003} 
has tended to be of quite a different vein to that
described here. In particular we employ a full likelihood analysis for
the fit to data in which the cosmological parameters are free to vary.
This freedom has important consequences for the independent use of the
temperature power spectrum to constrain such hybrid models --- at least
given the presently available data. Previous studies have also tended
to focus on local defects in the form of cosmic strings, a class of
models that we shall not look at here in any detail, and of such work only
\cite{Pogosian2003} has involved data from the WMAP project.

In our work, the cosmological perturbations are the uncorrelated sum
of those from (i) an inflationary adiabatic model, including both
scalar and tensor perturbations, and (ii) a global defect model, with
an $O(4)$ symmetry breaking at (or after) the end of inflation
producing textures (see e.g. \cite{Durrer2002}). In parametrizing the
primordial perturbations, we take there to be negligible running of
the scalar spectral index and that the tensor index obeys the
single-field consistency condition (see Sec. \ref{sec:MCMC}). We
also assume negligible neutrino masses, as would result from a
hierarchical mass variation with neutrino flavor. 

The contribution to the CMB power spectra from inflation is found
using a variant of the CMBFAST \cite{Seljak1996} approach, in the form
of CAMB \cite{Lewis2000}. The defect contribution is found by applying
the unequal-time correlator (UETC) approach \cite{Pen1997} to
numerical field evolution simulations, as will be discussed in the
next section. The UETC method addresses the problem that, in order to
calculate CMB power spectra to sub-degree scales, a simulation
conventionally requires a dynamic range that is far in excess of that
which is feasible with current technology. While it is possible to
make full-sky CMB maps \cite{Pen1994,Landriau2004}, this direct
approach is currently limited to relatively low multipoles: $\ell<20$,
although these potentially contain more information than the power
spectra alone. The UETC method boosts the dynamic range via a series
of theoretical simplifications, for example causality, such that high
resolution power spectra calculations may be performed from the
simulation data. To do so also involves calculations of the
CMBFAST-type, which we carried out using a version of CMBEASY
\cite{Doran2003} modified so as to deal with the defect scenario.

Despite the benefits of the UETC approach, the computational
requirements of CMB calculations that include defects far exceed those
that do not. As a result, the calculation of CMB spectra for a vast
number of different cosmological parameter values is not
attainable. Hence, the popular Markov chain Monte Carlo (MCMC)
approach \cite{Verde2003, Lewis2002, Gelman1992}, which involves many
thousand such calculations, cannot be fully applied to the defect
case. However, the non-defect case fits the WMAP data well and
defect-dominated structure formation does not give the required
acoustic peaks, suggesting that the defect contribution is small.  If
this is the case, then the result of a small change in the
cosmological parameters used for the defect calculation is a second
order effect. Therefore, the defect contribution needs only to be
calculated once, using currently favored values of the cosmological
parameters (see Sec. \ref{sec:CMBcalc}). The defect contribution is
then fixed, except for a normalization factor, which is free since it
is not known at which energy scale the defects formed. Hence the
approach used here, which is described more fully in Sec.
\ref{sec:MCMC}, is to apply the standard MCMC procedure to the
primordial contribution and add in the defect component with its
normalization varied as an MCMC parameter. This has been achieved
using a slightly modified version of CosmoMC \cite{Lewis2002}, which
is directly linked to CAMB. As this extra parameter controls the
degree to which the CMB power spectra differ from the usual non-defect
spectra, it shall be the main focus of this paper.

However, we shall not present our results in terms of this parameter
directly. Rather we shall use the fractional defect contribution to
the temperature power spectrum at a particular multipole,
$\ell=10$. The correspondence between the two is roughly linear for
low fractions but with a slight spread due to the variation of the
non-defect contribution to the chosen multipole. The fractional
quantity is, however, more directly understandable.

The data that we have used here are principally that from the first
year WMAP release \cite{Hinshaw2003, Kogut2003}: the temperature power
spectrum and the temperature-polarization (TE) cross-correlation
spectrum. Other CMB projects, such as ACBAR \cite{Kuo2004}, CBI
\cite{Pearson2003, Readhead2004} and VSA \cite{Dickinson2004}, which
give data out to higher multipoles than WMAP do not provide much in
the way of additional constraints on our model. Applying data on
cosmological parameters from, for example, work on big bang
nucleosynthesis (BBN) \cite{Kirkman2003} and measurements of the
Hubble parameter by the Hubble Key Project (HKP) \cite{Freedman2001}
has proved more important, as will be detailed in section
\ref{sec:MCMC}.

The changes to the matter power spectrum that the inclusion of defects
causes may be found in an entirely analogous manner to the CMB
calculation. However, while there have been recent
steps forward in measurements of this from galaxy redshift surveys,
such as 2DFGRS \cite{Percival2001} or SDSS \cite{Tegmark:2003ud},
and from the Lyman-$\alpha$ forest \cite{Croft2002}, we have
chosen not to use such data. Galaxy formation in the presence of
defect-induced density perturbations is not understood, and even in
pure inflation scenarios, inferences from the Lyman-$\alpha$ forest must
be drawn with care \cite{Seljak:2003jg}. Further, we do not believe that
the use of such data would significantly change our results, as we
shall discuss in Sec. \ref{sec:MCMC}. This is a conservative
position, driven by our desire to make reliable and statistically
meaningful statements about the relative importance of global defects.


\section{CMB calculations in global topological defect models}
\label{sec:CMBcalc}

The procedure to obtain the defect power spectra contribution for both
the CMB and for dark matter is as in Durrer, Kunz and Melchiorri
\cite{Durrer2002} and is fully detailed there. The method consists of
two distinct steps, the first of which is to compute the unequal time
two-point correlation functions $C_{\mu\nu\rho\lambda}$ of the defect energy
momentum tensor $T_{\mu\nu}$:
\begin{equation}
C_{\mu\nu\rho\lambda}(k,t,t') = 
\left\langle T_{\mu\nu}(k,t) T_{\rho\lambda}^{*}(k,t') \right\rangle .
\end{equation}
This is done with a numerical simulation of the classical $O(4)$
non-linear sigma model on a three-dimensional grid. For global
defects, where the seed energy momentum tensor can be taken to be
separately conserved, only five UETCs are independent, three for the
scalar perturbations and one each for the vector and tensor
perturbations. As topological defects generate perturbations at all
times after their creation, the vector perturbations do not decay and
have to be taken into account. Also, the relative amplitudes are fixed
by the model and cannot be adjusted. 

However, as already mentioned, the overall normalization of the
perturbations is related to be symmetry-breaking scale $\eta$ and is
free to be varied. Roughly, the relation to the CMB temperature
anisotropy $\delta T$ is as
\begin{equation} 
\frac{\delta T}{T} \sim \Psi \sim 4 \pi G \eta^{2},
\end{equation} 
where $\Psi$ is the dimensionless gravitational potential and $G$ is
Newton's constant. Hence, the dependence of $\eta$
upon the the normalization of power spectrum $\Ad$ (with the square
highlighting that the spectrum is quadratic in the perturbations) is
\begin{equation} 
\eta \sim (\Ad)^{\frac{1}{4}}.
\end{equation} 
Therefore $\Ad$ is not a sensitive measure of $\eta$. In fact, the
constraints upon $\eta$ from references herein are not likely to be
greatly changed by our results. Hence, here we focus solely
upon the significance of the defect perturbations compared to those of
primordial origin.

For this work, we used the simulations made for \cite{Durrer1999},
which used a $256^3$ grid. The results agree well with analytic
predictions \cite{Durrer1998} and the simulations of
\cite{Pen1997}. The UETCs were computed separately for radiation and
matter dominated backgrounds. In a background dominated by a
cosmological constant, the defects are quickly inflated away, so that
their contribution to the total energy-momentum tensor decays rapidly
and can be neglected.

The UETCs then act in the second step as external sources for a
Boltzmann solver. These codes need a deterministic source $S(k,t)$,
but the defects are essentially random by nature. We circumvent this
problem by diagonalizing each UETC, which through discretization and
the assumption of scaling evolution can be represented by a matrix with
indices $kt$ and $kt'$. This is hence a means of writing the full, 
incoherent source as the sum of coherent sources $v_n$ \cite{Turok1996}:
\begin{equation}
C(kt,kt') = \sum_n \lambda_n v_n(kt) v_n^*(kt') .
\end{equation}
To this end, we discretized the correlation functions into matrices of
size $200 \times 200$, which can then be diagonalized with the help of
standard methods. On a somewhat technical aside, the three scalar UETCs
are combined into one $400 \times 400$ matrix and diagonalized
together, as the third matrix represents the correlation (off-diagonal
part) between the other two.  The discretization can be performed in
different ways, e.g. by taking linear or logarithmic intervals in
$kt$. We use linear intervals as we found that this improves the
convergence of the results (but more care must be taken in this case
to ensure that the dynamical range is sufficient).

We interpolate the resulting eigenvectors $v_n(kt)$ with cubic
splines and use them as the sources for the Boltzmann solver. The
power spectra are then given by
\begin{equation}
C_\ell = \sum_n \lambda_n^{(S)} C_\ell^{(S)n} +
\sum_n \lambda_n^{(V)} C_\ell^{(V)n} +
\sum_n \lambda_n^{(T)} C_\ell^{(T)n} ,
\end{equation}
and correspondingly for the dark matter power spectrum $P(k)$. We use
the $20$ eigenvectors with the largest eigenvalues, which is more than
sufficient as the last ones contribute far less than 1\%.

Linear cosmological perturbation theory with seeds has been discussed
extensively in the literature. We work in the gauge-invariant
formalism of \cite{Durrer2002} with a modified version of the CMBEASY
Boltzmann code, using the total angular momentum method
\cite{Hu1997}. The sources are interpolated between matter and
radiation dominated epochs, and are gradually suppressed as the
cosmological constant starts to dominate.

\begin{figure}
\resizebox{\columnwidth}{!}{\includegraphics{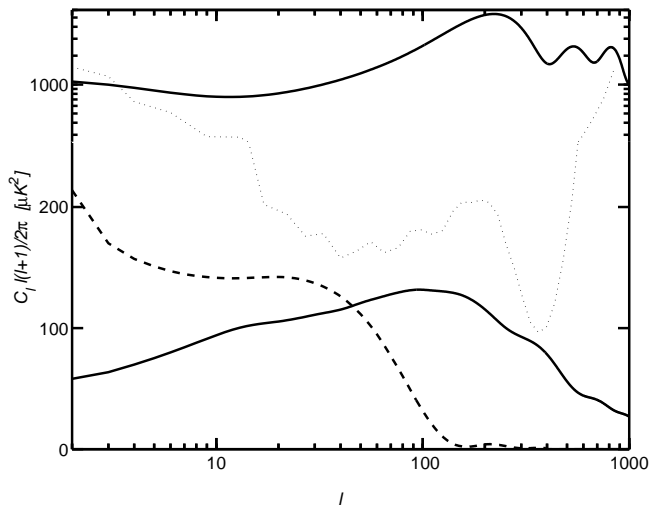}}
\caption{\label{fig:TT} The temperature power spectrum contributions
  from the global defects compared to that from primordial tensor
  perturbations and to the more dominant primordial scalar
  contribution. The defect and tensors contributions are scaled such
  that their contributions at $\ell=10$ are 13\% and 19\% respectively
  of a scalar-only fit to the WMAP data. (Note that the ordinate axis
  has linear scaling below $300\: \mu \rm K^2$ but logarithmic scaling above
  this value so as to show the slowly-varying defect contribution well
  on the same plot as the scalar contribution.) The dotted line
  indicates the 2$\sigma$ uncertainty in the WMAP data, including
  cosmic variance.}
\end{figure}

The fixed form of the defect contribution to the temperature power
spectrum for this case is illustrated in Fig.
\ref{fig:TT} (and was calculated using flat geometry, $h=0.70$,
$\Obhh=0.022$, $\Om=0.30$ and $\tau=0.10$ or $\zr=13$ ---  see Sec.
\ref{sec:MCMC} for definitions). The figure compares
it with the primordial scalar contribution (shown on a log scale) and
that resulting from primordial tensor perturbations. The defect
spectrum is scaled so as to match our final result for the 95\% upper
bound on the fractional contribution (see section \ref{sec:MCMC}). The
corresponding contribution to the TE cross-correlation spectrum is
very small and hence while it is incorporated in our calculations we
shall not illustrate it here (however, see \cite{Seljak1997} for a plot).

We expect that the overall error in the defect calculation is smaller
than about 10\%, and checked that our power spectra agree with other
published results \cite{Pen1997, Durrer2002} to within this accuracy.
Comparing the fixed defect spectrum used with a second one, calculated
using a different cosmology ($h=0.80$, $\Obhh=0.027$), shows that the
changes in the defect contribution are smaller than about 10\% over
the WMAP data range. For the case illustrated in Fig. \ref{fig:TT},
the defect contribution to the temperature power spectrum is of the
same order as the uncertainties from measurement plus cosmic variance,
in which case, such errors in our approach are not significant for the
comparison to data.

The final step is to simply add the defect contribution, with the
appropriate normalization, to that resulting from the primordial
scalar and tensor perturbations. This is justified by the fact that
the quantum fluctuations in the inflaton field and the symmetry
breaking field are uncorrelated during inflation when the global
symmetry is unbroken. Any subsequent interaction will be second order
in the gravitational perturbation and will hence be negligible.


\section{Monte Carlo parameter fitting}
\label{sec:MCMC}


\subsection{MCMC overview}

The MCMC approach has found recent application to CMB work principally
because it allows inferences to be made about the $n$ parameters in a
model without requiring a complete and detailed knowledge of the
$n$-dimensional likelihood surface. For example, to calculate the
likelihood surface on an $n$-dimensional grid consisting of $m$ points
in each dimension requires $m^{n}$ iterations. In the case of a four
parameter model, calculating 50 points in each direction requires $6\times 10^6$
iterations. Suppose that each iteration involves the
calculation of the CMB power spectra from primordial perturbations for
comparison to WMAP data.  This amounts to $3\: \rm s$ per iteration using CAMB
on a single processor of the UK National Cosmology Supercomputer
\footnote{We compile the April 2003 release of CAMB for the 1.3GHz
Intel Itanium II chips of this machine using the Intel Fortran
compiler version 7.}  and around 6 months in total. However, the MCMC
approach allows a reliable knowledge of the one-dimensional likelihood
function for each parameter (with all the other parameters integrated
out) in less than perhaps 100 000. This
corresponds to a few days of calculation and this is not sensitively
changed by the addition of extra parameters. Higher dimensional
functions may be found, in order to plot say a 95\% confidence
contours of parameter pairs, but the resolution is dependent upon the number of
iterations.

The MCMC method incorporated in CosmoMC is the Metropolis-Hastings
algorithm. This is a random walk through parameter space, with the
parameters recorded at each step to form a Markov chain. Each
iteration involves first the generation of a proposed next point in
the chain, found by generating a random change in one or more
parameters based upon some probability distribution. The form of the
distribution is not important for the end result, as long as the
probability of proposing point B from point A is the same as proposing
point A from point B. The power spectra are calculated for the
corresponding parameter set and then the likelihood of these spectra
is found from the data. If the likelihood at the proposed point is
greater than at the old point, then it is accepted and forms the next
point in the chain. Otherwise, a uniformly distributed random number
is generated between zero and unity and the proposed point is accepted
only if this number is greater than the ratio of the new to old
likelihoods. If the proposal is rejected, then the original point is
repeated in the chain.

In the limit of infinite such iterations, the number of points in
particular volume of parameter space is proportional to the
probability of the true parameters lying in that
volume. This has a number of attractive properties, for example, the
mean value of parameter in a chain is the expectation value of that
parameter given the data. Furthermore, the marginalized likelihood
function for a parameter is represented simply by a histogram of the
values from the chain.

Of course, an infinite chain is not attainable (our chains have
lengths of a few hundred thousand elements) and this presents a number
of problems. First, the chain is started from some arbitrary
location and the first thousand or so points may be highly dependent
upon this and carry a finite weighting in the chain. The usual
solution is to discard the beginning of the chain and we have followed
Gelman and Rubin \cite{Gelman1992} in removing the entire first
half. However, an analysis is still required to check if an inference
from the remaining points in the chain is reliable, given that a
finite chain may not have explored all of the relevant parameter space
to the same extent. The solution is to use $M$ chains (here we use
$M=5$), each started from a different point. If the results of all of
these chains agree, then confidence is high in any inferences made. To
decide if the $M$ chains compare favorably we have again followed the
approach of Gelman and Rubin. This is to make to a comparison for each
parameter between the spread within each chain and the spread of the
means from the chains. See Refs. \cite{Gelman1992} and
\cite{Verde2003} for a more details on this comparison.

While we have said that the form of the proposal distribution does not
affect the final result, it has a large effect on the efficiency of
the approach. The taking of very small steps requires a great many of
them to be made in order to fully explore the parameter space. On the
other hand, taking large steps increases the chance of
proposed-point rejections, and the exploration of the parameter space
is inefficient. The approach used here uses the covariance matrix,
found from a preliminary run, to set the appropriate step
scale. Furthermore, the direction of the step in the $n$-dimensional
space is determined using the covariance matrix diagonalization
approach implemented in CosmoMC, which deals well with linear
degeneracies in the parameter space.


\subsection{Model parameters}

The cosmological parameters that we have chosen to vary are (i) the
Hubble constant ($100h\:\rm{km}\:\rm{s}^{-1}\:\rm{Mpc}^{-1}$), (ii) the physical 
baryon density $\Obhh$, (ii) the total matter density $\Omhh$ (via the
cold dark matter density $\Ochh$), and (iv) the optical depth $\tau$ 
from the surface of decoupling or rather the redshift of
quasi-instantaneous re-ionization $\zr$. We have assumed that the 
Universe is flat as appropriate for an inflationary model, and in any 
case a change in the curvature gives an almost identical effect as a 
change in $h$, via the geometric degeneracy \cite{Efstathiou1999}. 
The effect of allowing a small curvature can therefore be created by 
change in the Hubble parameter. We will, however, consider
constraining $h$ later and so the explicit assumption of zero
curvature will then be made.  

The primordial power spectrum for scalar perturbations, or more precisely the
comoving curvature perturbation, has been parametrized by 
(v) the normalization $\As$ and (vi) the spectral index
$\ns$. The normalization is set at a comoving wavevector $k_{0}$ of
0.01 Mpc$^{-1}$ and we assume negligible variation of the spectral
index with scale. The scalar power spectrum is hence given by
\begin{equation}
\Power{s}=\As \left( \frac{k}{k_{0}} \right)^{\ns-1}.
\end{equation}

A finite contribution to the primordial perturbations from
gravitational waves has been allowed for. As shown in Fig.
\ref{fig:TT}, the contribution these tensor perturbations make to the
temperature spectrum is to raise very large scales only and hence the
tilt in this spectrum must be very large to be detectable in this
sub-dominant component. If we assume that the effective mass of the
field involved in the global symmetry breaking is much greater than
the Hubble parameter during inflation, and the hybrid model involves
only this field plus the inflaton, then we may use the single-field
inflation consistency relation. Following \cite{Leach2002} we use the
primordial form of this relation, giving the tensor tilt in terms of
the ratio of the primordial tensor and scalar power spectra at the
pivot scale $k_{0}$. This assumption is, however, not important given the
current data, and hence we are justified in parametrizing primordial tensor
perturbation only by (vii) the normalization $\At$ (via the ratio $\At/\As$):
\begin{equation}
\Power{t}=\At \left( \frac{k}{k_{0}} \right)^{ - \frac{1}{8}
\frac{\At}{\As} }.
\end{equation}
The final parameter that we have varied is then (viii) the
normalization of the defect contribution to the CMB power spectra
$\Ad$, giving a total of 8 parameters.


\subsection{Degeneracies, additional data and results}

As mentioned in the Introduction, we shall not present our results in
terms of $\Ad$ but in terms of the fractional defect contribution to
the temperature power spectrum at $\ell=10$: $\fd$ (and likewise for
tensors: $\ft$). This particular multipole lies in a region where the
temperature power spectrum is relatively flat and has become a
conventional place to make contribution ratios. However, this value of
$\ell$ happens to be roughly where the fractional defect contribution
is greatest. The value at the first peak is approximately one tenth of
this, a fact that should be taken into account when interpreting our
results.

Naively, $\fd$ is tightly constrained by the WMAP data because the
contribution to the temperature power spectrum is greatest at scales
were the data is very precise (see Fig. \ref{fig:TT}). However, as
is often the case in CMB parameter fitting, $\fd$ is involved in a
degeneracy with four other parameters, such that an increase in its
value may be compensated for by changes in the others and a fit to
data maintained. This may be understood broadly as follows. The effect
that increasing $\Ad$ has on the temperature power spectrum is to the
raise the region $\ell<400$. This may be reversed by lowering the
normalization of the primordial scalar perturbations using $\As$ and
giving a slight tilt toward small scales using $\ns$. (Increasing
$\tau$ also has the general effect of reducing the temperature
anisotropies, although the influence of $\tau$ tails off for
$\ell\lesssim100$ and is hence less important here.) Unfortunately,
the temperature power spectrum is most sensitive to the decrease in
$\As$ at the first peak and, as a result, it is excessively
lowered. Further tilting the primordial scalar spectrum can raise the
first peak but at the expense of raising the high-$\ell$ region too
far. However, an increase in $\Obhh$ raises the first peak while
lowering the region around the second and third peaks and therefore,
combined with extra tilt, this achieves the desired effect. The fit
may then be further improved, for the increases in both $\ns$ and
$\Obhh$ raise the high $\ell$ side of the first peak more than the low
$\ell$ side (outweighing the opposite effect of the defects). This
may be countered by using the Hubble constant to move the peak to
slightly lower $\ell$ and give an overall result that, considering the
data, is almost indistinguishable from the original.

\begin{figure}
\resizebox{\columnwidth}{!}{\includegraphics{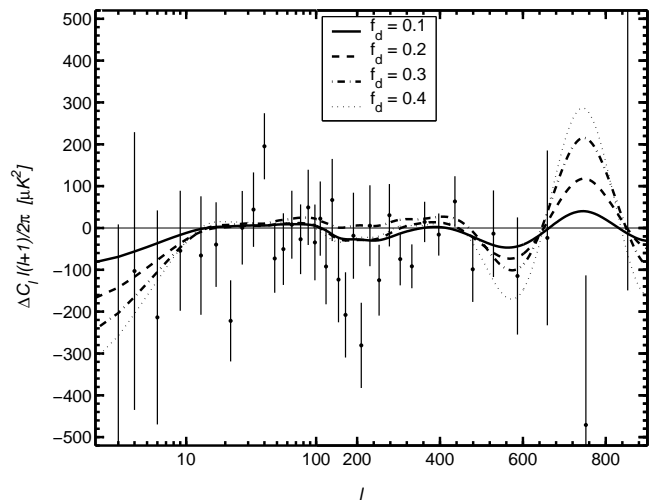}}
\caption{\label{fig:degen} The change in the temperature power
  spectrum from a model without defects, when defects are added and
  degeneracy direction followed to maintain the fit to data. The WMAP
  first-year binned data are over-plotted, with the zero-defect model
  subtracted from them, demonstrating the inability of the data to
  readily distinguish between the models. The error bars include
  cosmic variance. (Note that multipole axis is linear for $\ell>200$
  and logarithmic below this in order to show both regions clearly.)}
\end{figure}

\begin{table}
\begin{ruledtabular}
\begin{tabular}{ccccc}
$\fd$ & $\Obhh$ & $h$ & $10^{10}\As$ & $\ns$ \\
\hline
0.0 & 0.026 & 0.81 & 24 & 1.05 \\
0.1 & 0.028 & 0.86 & 22 & 1.10 \\
0.2 & 0.031 & 0.93 & 21 & 1.17 \\
0.3 & 0.034 & 1.03 & 19 & 1.25 \\
0.4 & 0.039 & 1.16 & 17 & 1.35 \\
\end{tabular}
\end{ruledtabular}
\caption{\label{tab:degen}Parameter values corresponding to the cases
considered in Fig. \ref{fig:degen} (see Sec. \ref{sec:MCMC} for
definitions). The following parameters are fixed: $\Omhh=0.13$,
$\zr=17$, $\ft=0.15$. The others are set as the means found from a
sub-set of an MCMC chain, selected such that the defect fraction
and these fixed parameters have approximately the correct values.}
\end{table}

\begin{figure}
\resizebox{\columnwidth}{!}{\includegraphics{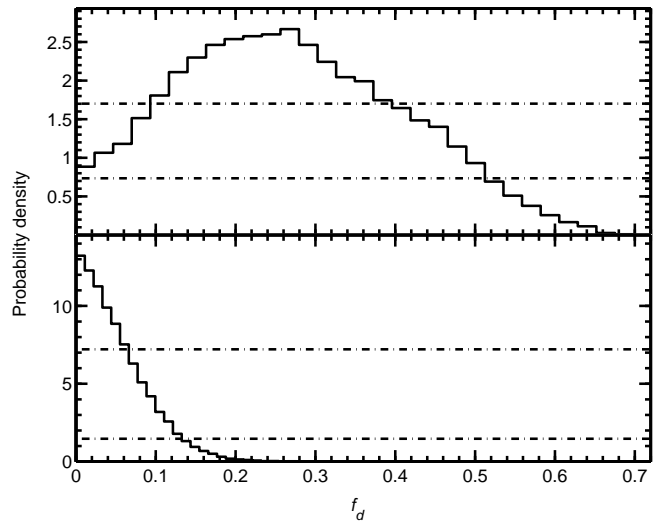}}
\caption{\label{fig:fd_hist} The marginalized likelihood function for
the defect fraction when using only the WMAP data (top) and when
additionally incorporating the BBN and HKP constraints (bottom). The
horizontal lines show the 68\% and 95\% confidence levels.}
\end{figure}

\begin{figure}
\resizebox{\columnwidth}{!}{\includegraphics{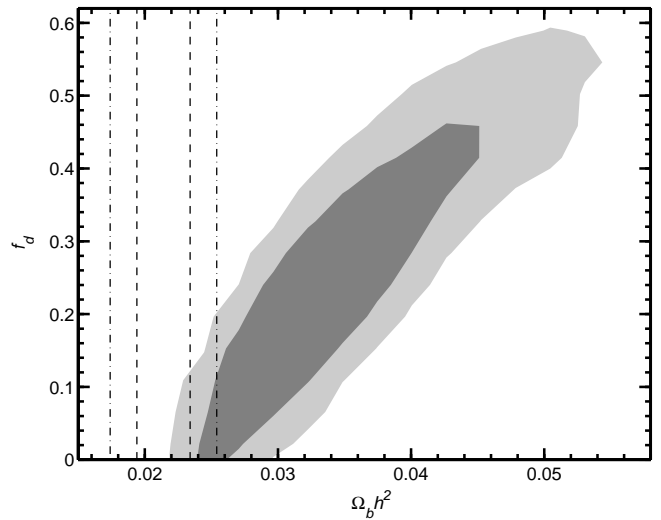}}
\caption{\label{fig:contour} A plot of the likelihood from the
  fit to WMAP data projected onto the \Obhh-\fd\ plane. The
  contours show the 68\% and 95\% confidence levels and highlight the
  degeneracy  between the two parameters. The vertical lines show the
  68\% and 95\% confidence limits of the determination of \Obhh of Kirkman \emph{et al}.}
\end{figure}

Figure \ref{fig:degen} shows the change in the temperature power
spectrum from a non-defect case that is caused by adding in defects at
four different levels: $\fd=0.1$, $0.2$, $0.3$, $0.4$. The other four
parameters of the degeneracy have been adjusted, as shown in Table
\ref{tab:degen}, so that the fit is maintained. Over-plotted are the
binned WMAP data with the non-defect model subtracted so that they may
be directly compared with the changes due to defects. The figure shows
that in the intermediate $\ell$ range, where the uncertainties are
small, the change in the power spectrum is minimal. However, the fit
on these scales cannot be preserved without deviations at small and
large scales, although these are not particularly large compared to
the uncertainties for the cases shown. As is evident from the
the marginalized likelihood for \fd\ shown in Fig.
\ref{fig:fd_hist}, the WMAP data allow for a substantial defect
contribution to the power spectrum. In fact, using the WMAP data
alone, a non-zero defect component of $\fd=0.27^{+0.13}_{-0.17}$ is preferred,
which amounts to a detection at around the 2$\sigma$ level. However, we do
not wish to claim that this is a significant result, firstly because
it relies upon unfavored values of the cosmological parameters. For
example, Fig. \ref{fig:contour} shows the degeneracy between \fd\ and
\Obhh. Also indicated is the BBN constraint that will be adopted
later, which differs greatly from the high \Obhh\ values necessary for
large defect fractions. But further, our numerical approach is
not readily capable of handling these large defect fractions, for the
method employed here is based upon an assumption to the
contrary. First, the numerical errors in the defect spectra
calculation are more relevant for large contributions. Second, if
the MCMC cosmology deviates too far from the fixed one for which the
defect spectrum was calculated (see Sec. \ref{sec:CMBcalc}) then it
is no longer applicable. Not only does the degeneracy mean that the
cosmology varies considerably, but the resulting inaccuracy is more
important at the higher fractions that it allows. Then finally, below
$\ell\sim350$ the WMAP uncertainties are dominated by cosmic variance
\cite{Hinshaw2003}, which is taken into account based upon Gaussian
statistics. Since the defect evolution is non-linear, they may
introduce a non-Gaussian component, which we require to be made
insignificant by the defect fraction being small. Thus, for both
scientific and practical reasons we do not wish to draw undue
attention to this result. We have merely turned to the use of
additional data in order to limit the effect of the degeneracy.

\begin{figure}
\resizebox{\columnwidth}{!}{\includegraphics{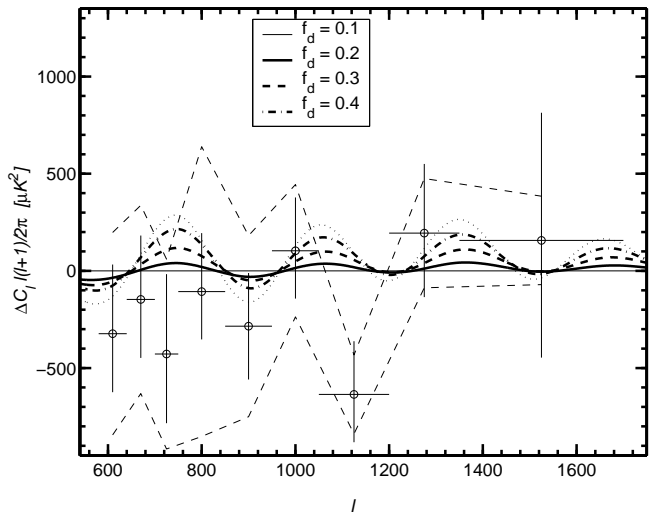}}
\caption{\label{fig:VSA} The changes to the temperature power spectrum
for the same cases as in Fig. \ref{fig:degen}, but shown at
higher multipoles and compared to the VSA data. The changes in the 
parameters follow the degeneracy present in the WMAP data but the 
changes in these previously unconstrained scales are too small for 
the VSA data to readily distinguish between the curves. The error bars 
do not include the 3\% calibration uncertainty, which allows the power
spectra values to be scaled up or down in unison. On this
difference plot, this uncertainty is shown by the dashed
zig-zags.}
\end{figure}

The use of other CMB data, on scales smaller than those probed by
WMAP is an obvious first step toward applying further constraint to
our model. Unfortunately, while each of ACBAR, CBI and VSA projects are
capable of sub-WMAP resolution, they do not have the precision to
limit the defect allowance to the point were the degeneracy is
no longer a problem. This is illustrated in Fig. \ref{fig:VSA} for the
VSA data. While of course their inclusion into our likelihood analysis
does provide additional constraint, our results are almost unchanged by
their addition and hence we shall not add unnecessary complication
by discussing these data further.

A more direct way of limiting the degeneracy is to constrain the
parameters $h$ and $\Obhh$. As noted above, in applying a constraint
on $h$ we are making our assumption of geometric flatness relevant and,
therefore, we shall first consider $\Obhh$. Big bang nucleosynthesis
calculations give predictions for the primordial abundances of the
light isotopes in terms of this parameter. However, it is not possible
to directly measure these, only the abundances in astronomical objects
at more recent times, and either assert that these abundances are
representative of the primordial era or make some allowance for
subsequent evolution. This is a notable problem and is perhaps
responsible for there being discrepancies between $\Obhh$
calculated using each of these isotopes. However, deuterium absorption
lines have been seen in the spectra of a number of quasars,
corresponding to the presence of gas clouds at high redshift. These
low (but finite) metallicity clouds are believed to have experienced
negligible deuterium processing and indeed show no notable correlation
between metallicity and deuterium abundance \cite{Kirkman2003,
Pettini2001}. Unfortunately, the number of such clouds that have been
well observed remains small and a number of difficulties exist in
extracting the D/H ratio. Indeed, the spread in the measurements is a
little large considering the estimated errors and Kirkman \emph{et
al.} \cite{Kirkman2003} conclude that it most likely as a result of
uncertainty under-estimation rather than any real variation. They
proceed to take the weighted mean of the log(D/H) measurements from
five such systems but use the spread in the values to provide the
uncertainty estimate. Their result is a value for $\Obhh$ of
$0.0214\pm0.0020$, which we use here. However, it should be noted
that uncertainties also arise from the use of nuclear cross-section
data and that these can change this result slightly \cite{Cyburt2004}.

Assuming a Gaussian form for the uncertainty, it is straightforward
to add such data to the MCMC approach, for it simply adds a Gaussian
term that multiplies the likelihood calculated from the WMAP data. As
would be expected considering Fig. \ref{fig:contour}, its
application is successful in limiting the effect of the degeneracy;
however, the result of $\Obhh=0.0245\pm0.0014$ is still a little high.
This is due to the WMAP data favoring a non-zero defect fraction, but
once the BBN data is taken into account, the preference for defects is
removed. The marginalized likelihood function for \fd\ is now highly
skewed with its peak at the zero-defect limit. We therefore present
merely the 95\% confidence upper bound: $\fd<0.14$ for case of the
WMAP data with the BBN constraint. We estimate the MCMC uncertainty in
this figure using the standard deviation between the values from the 5
independent chains, reduced by $\sqrt{5-1}$, and find that this gives
$0.002$ in this case. The Hubble constant is constrained to
$h=0.80\pm0.08$, although a direct interpretation of this value as $h$
would assume flatness, while the results for rest of the parameters
are shown in Table \ref{tab:params}.

\begin{table}
\begin{ruledtabular}
\begin{tabular}{cccc}
        & BBN             & HKP               & BBN and HKP \\
\hline
$h$     & $0.80\pm0.08$   & $0.80\pm0.06$    & $0.76\pm0.05$ \\
$\Obhh$ & $0.0245\pm0.0014$ & $0.0267\pm0.0021$  & $0.0242\pm0.013$ \\
$\Omhh$ & $0.126\pm0.018$ & $0.140\pm0.015$   & $0.134\pm0.014$ \\
$\zr$   & $13\pm5$        & $15\pm5$          & $13\pm4$ \\
$10^{10}\As$ & $22\pm3$   & $23\pm3$          & $22\pm2$ \\
$\ns$   & $1.03\pm0.04$   & $1.07\pm0.05$     & $1.01\pm0.03$ \\
$\ft$   & $<0.26$         & $<0.22$           & $<0.19$ \\
$\fd$   & $<0.14$         & $<0.23$           & $<0.13$ \\
\end{tabular}
\end{ruledtabular}
\caption{\label{tab:params}Parameter values when constrained by the
BBN and/or HKP results in addition to the first-year WMAP data (see
Sec. \ref{sec:MCMC} for definitions). Uncertainties indicated are 
standard deviations except for $\ft$ and $\fd$ for which 95\% 
upper bounds are shown.}
\end{table}

If the assumption of flatness is made explicit and the Hubble Key
Project result of $h=0.72\pm0.08$ \cite{Freedman2001} is applied
instead of the BBN constraint, then the result is broadly the
same. However, while $h$ is slightly more constrained in this case,
$\Obhh$ is less so and the upper bound on the defect contribution is
rather higher, now: $\fd<0.22$ (95\% confidence, 0.003 MCMC
uncertainty). However, this case does provide a bound that is
independent of the BBN constraint, which we otherwise rely very
heavily upon. If the two constraints are applied together, then the
BBN data dominate and the result is essentially the same as when the
BBN constraint was applied alone: $\fd<0.13$ (95\% confidence, 0.004
MCMC uncertainty). However, a number of the other parameters are more
tightly constrained than in the previous case, most notably the tensor
fraction. The effect of these constraints on the marginalized
likelihood function for the defect fraction is shown in Fig.
\ref{fig:fd_hist}, with the final result being peaked at $\fd=0$. The
levels of defect and tensor contributions that correspond to the
individual 95\% upper bounds in this constrained case are those that
were illustrated in Fig. \ref{fig:TT}.

While both defects and tensors contribute most to large scales,
tensors do so almost exclusively for $\ell<100$ and so the effects of
the two contributions are quite different. For example, while tensors
suffer from a degeneracy similar to that for $\Ad$, in the tensor case
there is an additional coupling to $\Omhh$. Also there is a stronger
coupling to $h$, as highlighted by the tensor contribution being more
constrained once the HKP result is incorporated. However, as a result of there
being some similarity in the two contributions, having a large tensor
contribution does limit the degree to which defects are allowable and
vice versa. But as neither the defect or tensor contributions can be
negative, adding an additional degree of freedom by allowing defects,
and so giving a positive expectation value for $\Ad$, further
constrains $\At$, rather than allowing it greater freedom.  The
reverse is also true, such that if we disallow primordial tensors in
our model, then the defect allowance increases a little. The
percentage increase in $\fd$ is about 15\% in the two cases involving the
BBN constrain and about 10\% in the case when the HKP constraint was applied
alone.

\begin{figure}
\resizebox{\columnwidth}{!}{\includegraphics{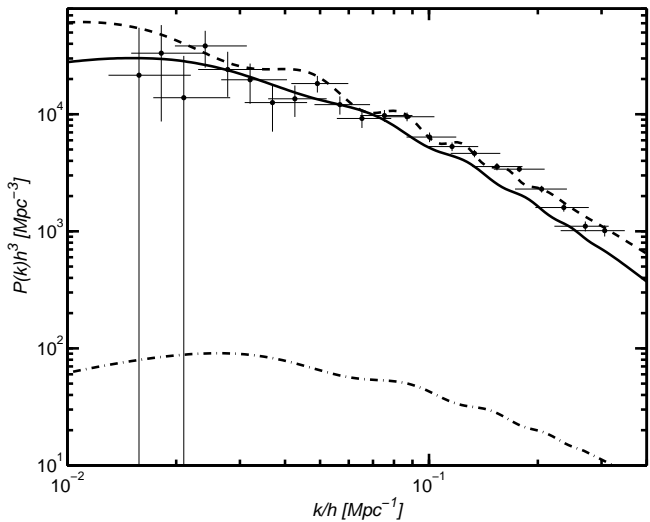}}
\caption{\label{fig:p_k} The current matter power spectrum in the
  non-defect case of Table \ref{tab:degen} (solid line) compared to
  that for the $\fd=0.4$ case (dashed line). The SDSS data is over
  plotted, showing an inability to distinguish the two cases. The
  sub-dominant defect contribution to the second case is also shown
  as the dot-dashed line.}
\end{figure}

As explained in the Introduction, we have not used the matter power
spectrum to constrain our model because of uncertainty about the
details of structure formation with global defects, caused by
the non-Gaussianity of the perturbations.  That non-Gaussianity can be
important was shown in Ref. \cite{Avelino2003}, where a model
of cosmic string-induced perturbations \cite{Wu:1998mr,Avelino1998}
was studied and estimated to contribute less than around 1\% of the
total matter power spectrum at $8h^{-1}$ Mpc.  The effect of the
non-Gaussianity was to make re-ionization earlier and slower than in
Gaussian models. While it is therefore inconclusive to simply compare
the matter spectra, it is still interesting. Figure \ref{fig:p_k}
compares one of the possible data sets, the SDSS data, with the
$\fd=0.0$ and $0.4$ cases from Table \ref{tab:degen}. The difference
between the two cases is partially due to the added defect
contribution, but is also due to the change in the cosmological
parameters required to maintain the fit to the WMAP data. It is
tempting to conclude that the SDSS data is not readily capable of
constraining the defect degeneracy and hence the limitation of the
degeneracy would again be dominated by the BBN constraint, but a
definitive conclusion requires much more work on the matter power
spectrum in models with global defects. Thus, we remain with our
conservative position with regard to the use of large-scale structure
data and have not incorporated it into our numerical results.


\section{Discussion and Conclusions}
\label{sec:con}

We have found that in order to constrain the extent to which global
defects assisted the seeding of cosmic structure, extra data in
addition to the CMB temperature power spectrum must be used, or we
require a greater knowledge of the spectrum than we have at
present. This is despite the uncertainties in the WMAP data set used
being dominated by cosmic-variance over the range where the defect
contribution to the power spectrum is greatest. The freedom in the
cosmological parameters and the primordial power spectrum are
sufficient to allow a significant defect contribution while still
fitting the data. It is only when parameters such as the physical
baryon density and the Hubble parameter are restricted, via
alternative astronomical techniques, that our model is significantly
constrained. Then we find that the temperature power spectrum at a
multipole of $\ell=10$ may have a fraction of 0.13 attributed to
defects (95\% confidence). The uncertainty in this upper bound that
comes from the MCMC approach is 3\%. More importantly, the numerical
errors in the defect calculation are believed to be of order 10\%, in
which case there may be of order 10\% change in this result. Also, if
the primordial tensor component was removed then this result would
increase by about 15\%. In addition, this bound is quite sensitive to
BBN result for the value for the physical baryon fraction $\Obhh$ used
and there has yet to be full agreement in this value among authors. 

A further result of the degeneracy found is that, if defects were to
contribute to cosmic structure formation, then there would a change in
the values of the cosmological parameters estimated from the current
CMB data. Most notably this affects $\Obhh$, $h$ and $\ns$ all of
which are subject to an increase upon the addition of defects. This acts
to re-enforce the more general point that any inferences made about
the cosmological parameters from the WMAP data are model dependent and
should be treated with caution.

We note that only a single defect type and model has been used in this
investigation: textures resulting from the breaking of an $O(4)$
symmetry. However, the contributions from other global defect models
are broadly the same as that considered here, although their spectra
are by no means identical. Considering the $O(N)$ class of models for
$N=2$ (strings), 3 (monopoles), 4 (textures) and 5 (non-topological
textures) there is a gradual variation with $N$ of the relative
contributions at low multipoles ($\ell\sim10$) and high multipoles
($\ell\sim300$), with strings giving preference to the latter
\cite{Pen1997}. This is likely to reduce the fractional contribution
allowed from strings by perhaps 30\% at $\ell=10$. Local defects in
the form of cosmic strings may give a broad peak at $\ell\sim400$
\cite{Battye2000, Contaldi1999}, beyond the first acoustic peak, and
hence the results may be quite different.

Further, we wish to point out that the WMAP project is ongoing and
new data with reduced uncertainties, as well as the addition of the EE
polarization power spectrum, will be released. While the defect
contribution to the EE spectrum is likely to be significant at only at
large scales, the freedom for defects when using the CMB power
spectra alone will be lessened by these data. It may be, however, that
the Planck satellite \cite{planck} is required to really explore
sub-dominant defect contributions using these power spectra
independently of other data. This mission will give precise
temperature power spectrum measurement at sub-WMAP resolutions. It
will also detect the B-mode polarization, which would be produced by
the vector and tensor components of the defect perturbations. Finally,
the non-Gaussianity of defect-induced perturbations has not been
satisfactorily addressed, either in the matter power spectrum or the
CMB. This may lead to more sensitive tests of defect scenarios.


\begin{acknowledgments}
N.B. and M.K. are supported by PPARC. We wish to thank Michael Doran for help
with the modification of CMBEASY and acknowledge extensive use
of the UK National Cosmology Supercomputer funded by PPARC, HEFCE and
Silicon Graphics.
\end{acknowledgments}


\bibliography{references}

\end{document}